\documentclass[12pt]{article}

\usepackage{cite,hyperref,amsfonts,amsmath,amssymb,bm,wasysym,a4wide}

\begin{document}

\title{\textbf{Chiral magnetic effect in the presence of electroweak interactions
as a quasiclassical phenomenon}}

\author{Maxim Dvornikov$^{a,b}$\thanks{maxdvo@izmiran.ru},\ 
Victor B. Semikoz$^{a}$\thanks{semikoz@yandex.ru}
\\
$^{a}$\small{\ Pushkov Institute of Terrestrial Magnetism, Ionosphere} \\
\small{and Radiowave Propagation (IZMIRAN),} \\
\small{108840 Moscow, Troitsk, Russia;} \\
$^{b}$\small{\ Physics Faculty, National Research Tomsk State University,} \\
\small{36 Lenin Avenue, 634050 Tomsk, Russia}}

\date{}

\maketitle

\begin{abstract}
We elaborate the quasiclassical approach to obtain the modified chiral magnetic effect
in the case when massless charged fermions interact with electromagnetic
fields and the background matter by the electroweak forces. 
The derivation of the anomalous current along the external magnetic field involves the study of the energy density evolution of chiral particles in parallel electric and magnetic fields. We consider both the particle acceleration by the external electric field and the contribution of the Adler anomaly. The condition of the validity of this method for the derivation of the chiral magnetic effect is formulated. We obtain the
expression for the electric current along the external magnetic field, which
appears to coincide with our previous results based on the purely
quantum approach. Our results are compared with findings of other
authors. 
\end{abstract}


%
%
%
%
%
%
%
%
%
%
%
%
%
%
%
%

\section{Introduction}

The chiral magnetic effect (CME), which consists in the generation
of the anomalous electric current flowing along the external magnetic
field, is widely applied in various branches of the modern physics. For example,
in accelerator physics this effect is used to account for some peculiarities
in heavy ion collisions. The applications of CME in astrophysics for
the explanation of strong magnetic fields in compact stars as well
as to account for great linear velocities of pulsars are known. CME
is also used in cosmology to consider the leptogenesis in strong cosmological
(hyper)magnetic fields. This effect is reported to be observed in 
such relativistic or pseudorelativistic systems as Dirac and Weyl
semimetals in condensed matter. The comprehensive references on the theory
and applications of CME are given, e.g., in Ref.~\cite{Kha15}.

Historically CME was theoretically predicted in Ref.~\cite{Vil80}
using the relativistic quantum mechanics approach based on the exact
solution of the Dirac equation for a charged particle in the external
magnetic field. Then, the result of Ref.~\cite{Vil80} was reproduced
in Ref.~\cite{NieNin83} basing on the study of the energy evolution of charged chiral particles in parallel electric and magnetic fields.

Recently, in Refs.~\cite{DvoSem15,DvoSem15b}, we generalized the results
of Ref.~\cite{Vil80} to account for the electroweak interaction
of charged fermions with background matter. We found that the expression
for the anomalous current gets the contribution from the parity violating
interaction. Although this contribution appears to be smaller than
the leading term in the current, it can be crucial on long time
intervals. For example, in Refs.~\cite{DvoSem15,DvoSem15b,Dvo16},
we proposed the application of this effect to explain the generation
of strong large scale magnetic fields in magnetars. The idea that the parity violating electroweak interaction between chiral particles can contribute to CME was also discussed in Ref.~\cite{BoyRucSha12}.

In this work, we present the new quasiclassical derivation of CME when chiral particles electroweakly interact with background matter. We also revisit the method for the description of CME, proposed in Ref.~\cite{NieNin83}, considering the particular case of the parity violating
electroweak interaction. We will focus on establishing the condition of the validity of the method of Ref.~\cite{NieNin83}, comparing it with the approach in Ref.~\cite{Vil80}. We will also consider the findings by other authors on this issue.

This work is organized as follows.
In Sec.~\ref{sec:DIREQ}, we remind how to find the energy spectrum of a massless fermion
moving in a background electroweak matter under the influence of the
external magnetic field. Then, in Sec.~\ref{sec:ANOMCURR}, we apply the method of Ref.~\cite{NieNin83} to derive the anomalous current. The evolution of the kinetic energy density is studied in Sec.~\ref{sec:EVKINEN}. We consider the contribution of the Adler anomaly to the evolution of the total energy density of the system of chiral particles in Sec.~\ref{sec:ADLER}. In Sec.~\ref{sec:COMP}, we compare different methods for the derivation of CME and find out when the approach of Ref.~\cite{NieNin83} is valid. The expression for the anomalous current is also obtained in Sec.~\ref{sec:COMP}. Finally, in Sec.~\ref{sec:DISSC} we discuss our results and compare them with the findings of other authors. The qualitative way to obtain CME is described in Appendix~\ref{sec:QUAL}.

\section{Solution of the Dirac equation\label{sec:DIREQ}}

Let us consider the system of charged fermions interacting by the electroweak forces in the Standard Model
with the background matter under the influence of an external electromagnetic
field $F_{\mu\nu}=(\mathbf{E},\mathbf{B})$. We shall suppose that
electric and magnetic fields are collinear and directed along the
$z$-axis: $\mathbf{E}=(0,0,\mathcal{E})$ and $\mathbf{B}=(0,0,\mathcal{B})$,
where $\mathcal{E}$ and $\mathcal{B}$ are constant values. Moreover,
we take that $\mathcal{B}\gg\mathcal{E}$. In this case we can treat
the electric field perturbatively. For the definiteness we shall assume
that the electric charge of a fermion (particle) is $q=-e<0$,
and an antifermion (antiparticle) has the electric charge $\bar{q}=e>0$.
Moreover the chiral symmetry is supposed to be unbroken, i.e. particles
are effectively massless. If the background matter is supposed to
be nonmoving and unpolarized, the Dirac equation for a fermion, described
by a bispinor $\psi$, has the form,
\begin{equation}\label{eq:Direq}
  \left[
    \gamma^{\mu}
    \left(
      \mathrm{i}\partial_{\mu}+eA_{\mu}
    \right) -
    \gamma^{0}
    \left(
      V_\mathrm{R}P_\mathrm{R}+V_\mathrm{L}P_\mathrm{L}
    \right)
  \right]
  \psi = 0,
\end{equation}
where $\gamma^{\mu}=\left(\gamma^{0},\bm{\gamma}\right)$ are the
Dirac matrices, $P_\mathrm{R,L}=(1\pm\gamma^{5})/2$ are the chiral projectors,
$\gamma^{5} = \mathrm{i}\gamma^{0}\gamma^{1}\gamma^{2}\gamma^{3}$, $A^{\mu}=(0,0,\mathcal{B}x,0)$
is the vector potential of the electromagnetic field, $V_\mathrm{R} \neq V_\mathrm{L}$ are
the effective potentials of the interaction of a fermion with background
matter. They are proportional to the densities of background fermions. The explicit form of $V_\mathrm{R,L}$ depends on the types of a
fermion and the background matter. For the electron-neutron interaction
these potentials are given in Refs.~\cite{DvoSem15,DvoSem15b}, and
for the quark-quark interaction in Refs.~\cite{Dvo15,Dvo16}. As mentioned
above, only the magnetic field will be taken into account exactly
in Eq.~(\ref{eq:Direq}).

The solution of Eq.~(\ref{eq:Direq}) was found in Refs.~\cite{DvoSem15,DvoSem15b}.
There are two independent solutions corresponding to different chiral
projections $\psi_\mathrm{R,L}(t)=P_\mathrm{R,L}\psi\sim\exp(-\mathrm{i}E_\mathrm{R,L}t)$. The
energy spectrum $E_\mathrm{R,L}$ for any of the chiral projection depends
on the discrete number $n=0,1,\dotsc$, which enumerates Landau levels.
For $n>0$ it has the form,
\begin{equation}\label{eq:En>0}
  E_\mathrm{R,L}=\pm V_\mathrm{R,L}+\sqrt{2e\mathcal{B}n+p_{z}^{2}},
\end{equation}
where $-\infty<p_{z}<+\infty$ is the eigenvalue of the projection
of the momentum operator along the $z$-axis $\hat{p}_{z}=-\mathrm{i}\partial_{z}$,
which commutes with the Hamiltonian of Eq.~(\ref{eq:Direq}). The
upper sign in Eq.~(\ref{eq:En>0}) corresponds to particles and the
lower one to antiparticles. If $n=0$, the energy spectra and the
allowed ranges of $p_{z}$ are summarized in Table~\ref{tab:Epzn=00003D0}.

\begin{table}
  \centering
  \protect
  \caption{The energy spectra and allowed values of $p_{z}$
  for fermions/antifermions at the lowest Landau level $n=0$
  as a result of the solution of Eq.~(\ref{eq:Direq}).
  \label{tab:Epzn=00003D0}}
  \begin{tabular}{lll}
    \hline 
    Type of particle/antiparticle & Energy spectrum & Allowed value of $p_{z}$   
    \tabularnewline
    \hline  
    Left fermion & $E_\mathrm{L}=V_\mathrm{L}+p_{z}$ & $p_{z}>0$
    \tabularnewline
    Right fermion & $E_\mathrm{R}=V_\mathrm{R}-p_{z}$ & $p_{z}<0$
    \tabularnewline
    Left antifermion & $E_\mathrm{L}=-V_\mathrm{L}-p_{z}$ & $p_{z}<0$
    \tabularnewline
    Right antifermion & $E_\mathrm{R}=-V_\mathrm{R}+p_{z}$ & $p_{z}>0$
    \tabularnewline
    \hline 
  \end{tabular}
\end{table}

\section{Anomalous current along the magnetic field\label{sec:ANOMCURR}}

In this section we shall apply the method for the description of CME proposed in Ref.~\cite{NieNin83} in the case when, besides the external magnetic field, massless charged particles electroweakly interact with background matter.

The derivation of the anomalous electric current of massless fermions in the parallel electric and magnetic fields is based on the evolution equation for the energy-momentum
tensor of matter $T_{\nu}^{\mu}$~\cite[p.~89]{LanLif88},
\begin{equation}\label{eq:emtevol}
  \partial_{\mu}T_{\nu}^{\mu}=F_{\nu\mu}J^{\mu},
\end{equation}
where $J^{\mu}=(J^{0},\mathbf{J})$ is the four vector of the current. Assuming that the fermionic matter is spatially homogeneous and considering the chosen configuration of the electromagnetic field, we get that the evolution of the total energy density of matter $\rho = T_0^0$ obeys the equation,
\begin{equation}\label{eq:emtrhom}
  \dot{\rho}=\mathcal{E}J_{z}.
\end{equation}
Here
\begin{equation}\label{eq:totendens}
  \rho = \sum_{\chi = \mathrm{R,L}}
  \left(
    E_\chi n_\chi + \bar{E}_\chi \bar{n}_\chi
  \right).
\end{equation}
where $E_\mathrm{R,L}$ and $\bar{E}_\mathrm{R,L}$ are the energies of a right and a left particle and antiparticle given in Eq.~\eqref{eq:En>0} and in Table~\ref{tab:Epzn=00003D0}, $n_\mathrm{R,L}$ and $\bar{n}_\mathrm{R,L}$ are the number densities of these fermions and antifermions.

Differentiating $\rho$ in Eq.~\eqref{eq:totendens} with respect to time and averaging over the statistical ensemble, we get that
\begin{equation}\label{eq:dotrhoav}
  \langle \dot{\rho} \rangle = \sum_{\chi = \mathrm{R,L}}
  \left\langle
    \dot{E}_{k\chi} n_\chi + \dot{\bar{E}}_{k\chi} \bar{n}_\chi +
    E_\chi \dot{n}_\chi + \bar{E}_\chi \dot{\bar{n}}_\chi
  \right\rangle.
\end{equation}
where $E_{k\mathrm{R,L}} = \bar{E}_{k\mathrm{R,L}} = \sqrt{2e\mathcal{B}n+p_{z}^{2}}$ are the kinetic energies of right and left particles and antiparticles. To derive Eq.~\eqref{eq:dotrhoav} we take into account that the kinetic energy of a charged particle changes when it interacts with an external electric field. Indeed, using the appropriate gauge,
we can modify the four potential of the considered electromagnetic
field to the form, $A^{\mu}=(-\mathcal{E}z,0,\mathcal{B}x,0)$. Then, basing on Eq.~\eqref{eq:Direq}, one can see that the operator $\hat{p}_{z}=-i\partial_{z}$ is no
longer an integral of Eq.~(\ref{eq:Direq}) since $A^{\mu}$ explicitly
depends on $z$. Hence $E_{k\mathrm{R,L}}=E_{k\mathrm{R,L}}(p_{z})$ is not conserved, i.e. the kinetic energy changes. The potential energies, $E_{p\mathrm{R,L}} = V_\mathrm{R,L}$ and $\bar{E}_{p\mathrm{R,L}} = - V_\mathrm{R,L}$, are constant by definition: $\dot{E}_{p\mathrm{R,L}} = \dot{\bar{E}}_{p\mathrm{R,L}} = 0$.

\subsection{Evolution of the kinetic energy density\label{sec:EVKINEN}}

To study the behavior of the kinetic energy, we start with the consideration of only left particles for simplicity. In the quasiclassical approximation, the evolution of the kinetic
energy of a particle $E_k$ obeys the equation~\cite[p.~51]{LanLif88},
\begin{equation}\label{eq:Ekevolclass}
  \dot{E}_k = q (\mathbf{E}\cdot \mathbf{v}),
\end{equation}
where $\mathbf{v}=\mathbf{p}/E_{k}$ is the particle velocity and $\mathbf{p}$ is its momentum. In our case, taking that $q = - e$ and $\mathbf{E} = \mathcal{E} \mathbf{e}_z$, we get that
\begin{equation}\label{eq:Ekevol}
  \left\langle
    \dot{E}_{k\mathrm{L}} n_\mathrm{L}
  \right\rangle  =
  -e\mathcal{E}
  \left\langle
    v_{z} n_\mathrm{L}
  \right\rangle.
\end{equation}
The average of the particle velocity in Eq.~\eqref{eq:Ekevol} can be computed using the phase volume in the external magnetic field~\cite[p.~173]{LanLif02},
\begin{equation}\label{eq:phvol}
  \left\langle
    v_{z} n_\mathrm{L}
  \right\rangle =
  \frac{e\mathcal{B}}{(2\pi)^{2}}
  \sum_{n=0}^\infty \int \mathrm{d}p_{z} v_z
  f(E_\mathrm{L}-\mu_\mathrm{L}),
\end{equation}
where $f(E)=\left[\exp(\beta E)+1\right]^{-1}$ is the Fermi-Dirac
distribution function, $\beta=1/T$ is the reciprocal temperature,
and $\mu_\mathrm{L}$ is the chemical potential of left particles. In Eq.~\eqref{eq:phvol},
we omit the factor $g_{s}=2$ since left particles have only one polarization;
cf. Ref.~\cite[p.~173]{LanLif02}. To derive Eq.~\eqref{eq:phvol} we imply the normalization of the distribution function,
\begin{equation}\label{eq:nLdef}
  n_\mathrm{L} =
  \frac{e\mathcal{B}}{(2\pi)^{2}}
  \sum_{n=0}^\infty \int \mathrm{d}p_{z}
  f(E_\mathrm{L}-\mu_\mathrm{L}),
\end{equation}
and analogous expressions for $n_\mathrm{R}$ and $\bar{n}_\mathrm{R,L}$.


Using the fact that $v_z = p_z /(E_\mathrm{L} - V_\mathrm{L})$ in the integrand in Eq.~\eqref{eq:phvol}, we get that 
\begin{equation}\label{eq:vav}
  \left\langle
    v_{z} n_\mathrm{L}
  \right\rangle =
  \frac{e\mathcal{B}}{4\pi^{2}}
  \sum_{n=0}^{\infty}
  \int \mathrm{d} p_{z}
  \frac{p_{z}}{E_\mathrm{L}-V_\mathrm{L}}
  f(E_\mathrm{L}-\mu_\mathrm{L}).
\end{equation}
%
With help of Eq.~(\ref{eq:En>0}), one gets that, at $n>0$, the integrand
in Eq.~(\ref{eq:vav}) is the odd function of $p_{z}$, making the
whole integral to vanish since the integration is in symmetric limits:
$-\infty<p_{z}<+\infty$. For $n=0$, taking into account Table~\ref{tab:Epzn=00003D0},
Eq.~(\ref{eq:vav}) can be rewritten in the form,
\begin{equation}
  \left\langle
    v_{z} n_\mathrm{L}
  \right\rangle =
  \frac{e\mathcal{B}}{4\pi^{2}}
  \int_{0}^{+\infty}\mathrm{d} p_{z}
  f(p_{z}+V_\mathrm{L}-\mu_\mathrm{L}).
\end{equation}
Therefore the contribution of left
particles to the evolution of the kinetic energy density of matter is
\begin{equation}\label{eq:phomL}
  \left\langle
    \dot{E}_{k\mathrm{L}} n_\mathrm{L}
  \right\rangle =
  -\frac{e^{2}\mathcal{EB}}{4\pi^{2}}
  \int_{0}^{+\infty}\mathrm{d} p_{z}
  f(p_{z}+V_\mathrm{L}-\mu_\mathrm{L}).
\end{equation}
If we include the contributions of left antifermions
as well as of right fermions and antifermions,
analogously to the derivation of Eq.~(\ref{eq:phomL}), we get that
the evolution of the total kinetic energy density
obeys the equation,
\begin{multline}\label{eq:rhomtot}
  \sum_{\chi = \mathrm{R,L}}
  \left\langle
    \dot{E}_{k\chi} n_\chi + \dot{\bar{E}}_{k\chi} \bar{n}_\chi
  \right\rangle
  \\
  =
  \frac{e^{2}\mathcal{EB}}{4\pi^{2}}
  \bigg\{
    \int_{-\infty}^{0}\mathrm{d} p_{z}
    f(-p_{z}+V_\mathrm{R}-\mu_\mathrm{R}) -
    \int_{0}^{+\infty}\mathrm{d} p_{z}
    f(p_{z}-V_\mathrm{R}+\mu_\mathrm{R})
    \\
    -
    \int_{0}^{+\infty}\mathrm{d} p_{z}
    f(p_{z}+V_\mathrm{L} -\mu_\mathrm{L}) +
    \int_{-\infty}^{0}\mathrm{d} p_{z}
    f(-p_{z}-V_\mathrm{L}+\mu_\mathrm{L})
  \bigg\}.
\end{multline}
where $\mu_\mathrm{R}$ is the chemical potential of right particles.

A remark on the signs in Eq.~(\ref{eq:rhomtot}) should be made.
The signs of the antiparticles contributions are opposite to that of
particles since the sign of the charge in Eq.~(\ref{eq:Ekevolclass})
is different for antiparticles. The chemical potentials of antiparticles
are taken with the opposite sign as usual. The arguments of the distribution
functions are based on Table~\ref{tab:Epzn=00003D0}. The sign of
the right fermions contribution is opposite to that of left fermions
since, in the integrand, one has $p_{z}/(E_\mathrm{R}-V_\mathrm{R})=-1$, whereas
$p_{z}/(E_\mathrm{L}-V_\mathrm{L})=+1$; cf. Table~\ref{tab:Epzn=00003D0}.

The direct computation of integrals in Eq.~(\ref{eq:rhomtot}) gives~\cite{DvoSem15,DvoSem15b}
\begin{equation}\label{eq:rhomm5v5}
  \sum_{\chi = \mathrm{R,L}}
  \left\langle
    \dot{E}_{k\chi} n_\chi + \dot{\bar{E}}_{k\chi} \bar{n}_\chi
  \right\rangle =
  \frac{e^{2}}{2\pi^{2}}(\mu_{5}+V_{5})\mathcal{EB},
\end{equation}
where $\mu_{5}=(\mu_\mathrm{R}-\mu_\mathrm{L})/2$ and $V_{5}=(V_\mathrm{L}-V_\mathrm{R})/2$.
Note that Eq.~(\ref{eq:rhomm5v5}) is independent of the plasma temperature.

The evolution of the kinetic energy density in Eq.~\eqref{eq:rhomm5v5} is based on the classical electrodynamics expression for the change of the particle kinetic energy in Eq.~\eqref{eq:Ekevolclass}. The quasiclassical approximation is known to be valid at $n \gg 1$~\cite{Coh77}. As results from Eq.~\eqref{eq:rhomtot}, the lowest Landau level yields the main contribution to the kinetic energy density evolution. The wave function of a particle at $n=0$ depends on both $z$ and the transversal coordinates $\mathbf{r}_\perp \in (x,y)$. Therefore quantum fluctuations of the particle velocity in the transverse plane are possible: $\mathbf{v}_\perp^\mathrm{(quant)} \neq 0$. However, taking into account that the electric field is along the $z$-axis, $\mathbf{E} = \mathcal{E} \mathbf{e}_z$, we get that these quantum fluctuations are irrelevant in our case since $(\mathbf{v}_\perp^\mathrm{(quant)} \cdot \mathbf{E}) = 0$ in Eq.~\eqref{eq:Ekevolclass}. At $n=0$, the particle motion along the $z$-axis is not affected by the magnetic field, and hence, can be treated as classical. Thus the quasiclassical approximation is valid for the description of the kinetic energy density evolution in Eq.~\eqref{eq:rhomtot}.

We can provide a different derivation of Eq.~\eqref{eq:rhomm5v5}. Again, let us start with the consideration of only left particles. If a constant electric field is applied along the $z$-axis, a fermion, having the charge $q = -e$, acquires the momentum along the $z$-axis $\mathrm{d}p_z = - e \mathcal{E} \mathrm{d}t$ during the time interval $\mathrm{d}t$. The change of the kinetic energy density of left particles, averaged over the statistical ensemble, $\left\langle n_\mathrm{L}\mathrm{d}E_{k\mathrm{L}} \right\rangle = \left\langle n_\mathrm{L} E_{k\mathrm{L}} \right\rangle|_{t+\mathrm{d}t} - \left\langle n_\mathrm{L} E_{k\mathrm{L}} \right\rangle|_{t}$, is
\begin{align}\label{eq:Ekchange}
  \left\langle
    n_\mathrm{L}\mathrm{d}E_{k\mathrm{L}} 
  \right\rangle = &
  \frac{e\mathcal{B}}{(2\pi)^2}\sum_{n=0}^{\infty}
  \int \mathrm{d}p_{z} (E_\mathrm{L}-V_\mathrm{L})
  \nonumber
  \displaybreak[1]
  \\
  & \times
  \left[
    f(E_\mathrm{L}(p_{z}-\mathrm{d}p_{z})-\mu_\mathrm{L}) -
    f(E_\mathrm{L}(p_{z})-\mu_\mathrm{L})
  \right]
  \notag
  \displaybreak[1]
  \\
  & \approx
  \frac{e^2\mathcal{EB}}{(2\pi)^2}\mathrm{d} t
  \sum_{n=0}^{\infty} \int \mathrm{d}p_{z}
  (E_\mathrm{L}(p_z)-V_\mathrm{L})
  \left.
    \frac{\mathrm{d}f}{\mathrm{d}p_z}
  \right|_{E_\mathrm{L}(p_{z})-\mu_\mathrm{L}},
\end{align}
where, again, we use the phase space volume in a magnetic field~\cite[p.~173]{LanLif02}. In fact, Eq.~\eqref{eq:Ekchange} is analogous to the application of a classical kinetic equation for the distribution functions of chiral particles to describe the change of the kinetic energy. The action of the magnetic field is taken into account in Eq.~\eqref{eq:Ekchange} by summing over the Landau levels rather than by accounting for the Lorentz force.

As results from Eq.~\eqref{eq:En>0}, at $n>0$, the energy levels are even functions of $p_z$ whereas $\mathrm{d}f/\mathrm{d}p_z$ is the odd function. Hence the contribution of higher Landau levels to the change of the kinetic energy vanishes since the integration is in symmetric limits: $-\infty < p_z < + \infty$. Thus, only the lowest Landau level with $n=0$ contributes to Eq.~(\ref{eq:Ekchange}) giving one
\begin{align}\label{eq:Edotpart}
  \left\langle
    n_\mathrm{L} \dot{E}_{k\mathrm{L}}
  \right\rangle = &
  \frac{e^{2}\mathcal{EB}}{4\pi^{2}}\int_{0}^{+\infty}\mathrm{d}p_{z}\, p_{z}
  \left. 
    \frac{\mathrm{d}f}{\mathrm{d}p_{z}}
  \right|_{p_{z}+V_\mathrm{L}-\mu_\mathrm{L}}
  \notag
  \\
  & =
  -\frac{e^{2}\mathcal{EB}}{4\pi^{2}}
  \int_{V_\mathrm{L}-\mu_\mathrm{L}}^{+\infty}\mathrm{d}pf(p),
\end{align}
where we integrate by parts and use the fact that $f(+\infty)=0$. Note that, to get Eq.~\eqref{eq:Edotpart}, it is important that $0<p_z<+\infty$ for left electrons; cf. Table~\ref{tab:Epzn=00003D0}.

The contribution of left antiparticles to the evolution of the kinetic
energy density $\left\langle \bar{n}_{\mathrm{L}} \dot{\bar{E}}_{k\mathrm{L}}\right\rangle $ can be obtained
from Eq.~(\ref{eq:Edotpart}) by changing the total sign of the expression as well
as replacing $V_\mathrm{L}\to-V_\mathrm{L}$ and $\mu_\mathrm{L}\to-\mu_\mathrm{L}$. Eventually
one gets,
\begin{align}\label{eq:ELdot}
  \left\langle
    n_\mathrm{L} \dot{E}_{k\mathrm{L}} +
    \bar{n}_\mathrm{L} \dot{\bar{E}}_{k\mathrm{L}} 
  \right\rangle  = &
  \frac{e^{2}\mathcal{EB}}{4\pi^{2}}
  \left[
    \int_{\mu_\mathrm{L} -V_\mathrm{L}}^{+\infty}\mathrm{d}pf(p) -
    \int_{V_\mathrm{L}-\mu_\mathrm{L}}^{+\infty}\mathrm{d}pf(p)
  \right]
  \nonumber
  \\
  & =
  -\frac{e^{2}}{4\pi^{2}}(\mu_\mathrm{L}-V_\mathrm{L})\mathcal{EB}.
\end{align}
Note that Eq.~\eqref{eq:ELdot} is valid for any temperature of the fermion gas.

The contribution of right particles and antiparticles to the kinetic
energy density evolution is derived analogously to Eq.~(\ref{eq:ELdot}):
one should change the total sign as well as replace $V_\mathrm{L}\to V_\mathrm{R}$
and $\mu_\mathrm{L}\to\mu_\mathrm{R}$. The final result, including all the cotributions, 
reads
\begin{equation}\label{eq:EdotRL}
  \sum_{\chi = \mathrm{R,L}}
  \left\langle
    n_\chi \dot{E}_{k\chi} + \bar{n}_\chi \dot{\bar{E}}_{k\chi}
  \right\rangle =
  \frac{e^{2}}{2\pi^{2}}(\mu_{5}+V_{5})\mathcal{EB},
\end{equation}
which completely coincides with Eq.~\eqref{eq:rhomm5v5}.


\subsection{Contribution of the Adler anomaly\label{sec:ADLER}}

Besides the terms describing the acceleration of chiral particles in parallel electric and magnetic fields, which were studied in Sec.~\ref{sec:EVKINEN}, in Eq.~\eqref{eq:dotrhoav}, there are contributions proportional to the change of the particle and antiparticle densities in this electromagnetic field. These terms are related to the Adler anomaly for massless charged particles.

The Adler anomaly reads~\cite{PesSch95}
\begin{equation}\label{eq:Aa}
  \partial_\mu
  \left(
    j_\mathrm{R}^\mu - j_\mathrm{L}^\mu
  \right) = 
  \frac{e^2}{2\pi^2}
  \left(
    \mathbf{E} \cdot \mathbf{B}
  \right),
\end{equation}
where $j_\mathrm{R,L}^\mu = \langle \bar{\psi}_\mathrm{R,L} \gamma^\mu \psi_\mathrm{R,L} \rangle$ are the four currents of chiral particles. Integrating Eq.~\eqref{eq:Aa} over the whole space and assuming that the three currents are vanishing at the infinity, we get that
\begin{equation}\label{eq:Aan}
  \frac{\mathrm{d}}{\mathrm{d}t}
  \left(
    n_\mathrm{R} - n_\mathrm{L} - \bar{n}_\mathrm{R} + \bar{n}_\mathrm{L}
  \right) = 
  \frac{e^2}{2\pi^2}
  \mathcal{EB}.
\end{equation}
To derive Eq.~\eqref{eq:Aan} we take that $j^0_\mathrm{R,L} = n_\mathrm{R,L} - \bar{n}_\mathrm{R,L}$. Comparing Eqs.~\eqref{eq:rhomm5v5} and~\eqref{eq:Aan}, we can see that the contribution of the Adler anomaly to $\langle \dot{\rho} \rangle$ can be comparable with that in Eq.~\eqref{eq:rhomm5v5}. Therefore one should carefully estimate the magnitude of the last two terms in Eq.~\eqref{eq:dotrhoav}.

Again we shall start with the studies of only left particles for simplicity. Supposing that $\langle \dot{n}_\mathrm{L} E_\mathrm{L} \rangle = \dot{n}_\mathrm{L} \langle E_\mathrm{L} \rangle$, we get that
\begin{equation}\label{eq:dotEdef}
  \langle E_\mathrm{L} \dot{n}_\mathrm{L} \rangle =
  \frac{\dot{n}_\mathrm{L}}{n_\mathrm{L}}
    \frac{e\mathcal{B}}{(2\pi)^{2}}
  \sum_{n=0}^\infty \int \mathrm{d}p_{z} E_\mathrm{L}
  f(E_\mathrm{L}-\mu_\mathrm{L}),
\end{equation}
where $n_\mathrm{L}$ is defined in Eq.~\eqref{eq:nLdef}. In the general situation, the integrals in Eq.~\eqref{eq:dotEdef} depend on $\mu_\mathrm{L}$, $T$, and $\mathcal{B}$. We will be mainly interested in the analysis of a strongly degenerate matter, for which $(\mu_\mathrm{L} - V_\mathrm{L}) \gg T$. In this situation, we can disregard the contribution of antiparticles.

For the illustration, let us discuss two opposite cases of (i) an extremely strong and (ii) a weak magnetic fields. First, let us study the case (i) supposing that $\mathcal{B} > (\mu_\mathrm{L} - V_\mathrm{L})^2/2e$. In this situation, all particles were found in Ref.~\cite{Nun97} to occupy only the lowest energy level with $n=0$. Computing the integrals over $p_z$ using the results of Ref.~\cite[pp.~169--170]{LanLif02}, we get that Eq.~\eqref{eq:dotEdef} takes the form,
%
\begin{equation}\label{eq:dotEL}
  \langle E_\mathrm{L} \dot{n}_\mathrm{L} \rangle =
  \frac{\dot{n}_\mathrm{L}}{2}
  \left[
    \mu_\mathrm{L} + V_\mathrm{L} +
    \frac{\pi^2 T^2}{3(\mu_\mathrm{L} - V_\mathrm{L})}
  \right].
\end{equation}
Taking into account for the contribution of right particles, which is analogous to Eq.~\eqref{eq:dotEL}, we can obtain that
\begin{equation}\label{eq:dotnRL}
  \sum_{\chi = \mathrm{R,L}}
  \left\langle
    E_\chi \dot{n}_\chi
  \right\rangle = 
  \frac{e^2}{4\pi^2}\mathcal{EB}
  \left[
    \mu_5 - V_5 +
    \frac{\pi^2 T^2}{6}
    \left(
      \frac{1}{\mu_\mathrm{R} - V_\mathrm{R}} -
      \frac{1}{\mu_\mathrm{L} - V_\mathrm{L}}
    \right)
  \right].
\end{equation}
To derive Eq.~\eqref{eq:dotnRL} we use Eq.~\eqref{eq:Aan} and suppose that the total number of particles is constant: $\dot{n}_\mathrm{R} + \dot{n}_\mathrm{L} = 0$.

In the opposite case (ii), when $\mathcal{B} \ll (\mu_\mathrm{L} - V_\mathrm{L})^2/2e$, multiple Landau levels contribute to the integrals in Eq.~\eqref{eq:dotEdef}. The analogue of Eq.~\eqref{eq:dotnRL} in the situation of a rather weak magnetic field reads
\begin{equation}\label{eq:dotnRLweak}
  \sum_{\chi = \mathrm{R,L}}
  \left\langle
    E_\chi \dot{n}_\chi
  \right\rangle \approx 
  \frac{e^2}{8\pi^2}\mathcal{EB}
  \left[
    3\mu_5 - V_5 +
    \frac{3\pi^2 T^2}{2}
    \left(
      \frac{1}{\mu_\mathrm{R} - V_\mathrm{R}} -
      \frac{1}{\mu_\mathrm{L} - V_\mathrm{L}}
    \right)
  \right],
\end{equation}
where we neglect terms $\sim T^4$.

In the general situation of the arbitrary temperature and chemical potentials, the terms in $\langle \dot{\rho} \rangle$ in Eq.~\eqref{eq:dotrhoav} containing the time derivative of number densities of right and left particles and antiparticles cannot be transformed to the simple forms like in Eqs.~\eqref{eq:dotnRL} and~\eqref{eq:dotnRLweak}. The problem is that the two conditions --- the Adler anomaly in Eq.~\eqref{eq:Aa} and the conservation of the total number of particles and antiparticles, $\dot{n}_\mathrm{R} + \dot{n}_\mathrm{L} + \dot{\bar{n}}_\mathrm{R} + \dot{\bar{n}}_\mathrm{L}= 0$ --- are not sufficient to determine the derivatives of the four number densities, $\dot{n}_\mathrm{R,L}$ and $\dot{\bar{n}}_\mathrm{R,L}$.
%

\subsection{Comparison of different approaches for the derivation of CME\label{sec:COMP}}

The first derivation of anomalous current of massless particles along the external magnetic field was performed in Ref.~\cite{Vil80}. This derivation is based on the direct calculation of the current using the exact solution of the Dirac equation for a charged particle in an external magnetic field. The result of Ref.~\cite{Vil80} reads
\begin{equation}\label{eq:J5Vil}
  \mathbf{J} =  \frac{e^{2}}{2\pi^{2}}\mu_{5}\mathbf{B},
\end{equation}
Recently, in Refs.~\cite{DvoSem15,DvoSem15b}, we generalized this finding of Ref.~\cite{Vil80} to include the electroweak interaction of charged particles with background matter. It leads to the replacement $\mu_{5} \to \mu_{5} + V_{5}$ in Eq.~\eqref{eq:J5Vil}.

Then, the alternative derivation of CME, based on the energy density evolution of massless charged particles in parallel electric and magnetic fields, was proposed in Ref.~\cite{NieNin83}. Formally, the expression for the current, obtained in Ref.~\cite{NieNin83}, coincides with that in Eq.~\eqref{eq:J5Vil}. However, the fact that the number densities of massless particles can be not constant gives an additional contribution to the total energy density evolution, as seen in Eq.~\eqref{eq:dotrhoav}. The non-conservation of the number densities of massless charged particles in parallel electric and magnetic fields is driven by the Adler anomaly in Eq.~\eqref{eq:Aa}. We have shown in Sec.~\ref{sec:ADLER} that this additional contribution to the energy density evolution strongly depends on the chemical potentials, the temperature, and the magnetic field strength; cf. Eqs.~\eqref{eq:dotnRL} and~\eqref{eq:dotnRLweak}.

Therefore, to reach the agreement between the two methods for the derivation of CME in Refs.~\cite{Vil80} and~\cite{NieNin83}, even at the absence of the electroweak background matter, i.e. when $V_\mathrm{R} = V_\mathrm{L} = 0$, we should neglect the contribution of the Adler anomaly in the method in Ref.~\cite{NieNin83}. It means that the number densities of right and left particles and antiparticles should be considered constant. The action of the external electric field consists just in the acceleration of these charged particles within each particle species. Thus we should set $\dot{n}_\mathrm{R,L} = \dot{\bar{n}}_\mathrm{R,L} = 0$ in Eq.~\eqref{eq:dotrhoav}. It means that massless charged particles should be considered as classical, with other quantum effects, like the Adler anomaly, being disregarded. Only in this case the methods in Refs.~\cite{Vil80} and~\cite{NieNin83} give the coinciding expressions for the anomalous current in Eq.~\eqref{eq:J5Vil}.

Now, let us consider the interaction of massless charged particles with an electroweak background matter and apply the analogue of the method in Ref.~\cite{NieNin83} for the derivation of the anomalous current. Combining Eq.~\eqref{eq:dotrhoav}, where $\dot{n}_\mathrm{R,L} = \dot{\bar{n}}_\mathrm{R,L} = 0$, with Eqs.~\eqref{eq:emtrhom} and~\eqref{eq:rhomm5v5} we get that the current along the magnetic field becomes,
\begin{equation}\label{eq:J5DvoSem}
  \mathbf{J} =  \frac{e^{2}}{2\pi^{2}}(\mu_{5} + V_5)\mathbf{B},
\end{equation}
where we restore the vector notations. Note that the expression for $\mathbf{J}$ in Eq.~\eqref{eq:J5DvoSem} completely coincides with that obtained in Refs.~\cite{DvoSem15,DvoSem15b}.

\section{Discussion\label{sec:DISSC}}

In the present work, the quasiclassical
derivation of CME in the presence of the electroweak interaction of
charged fermions with background matter was carried out. In our analysis
we followed the general idea of Ref.~\cite{NieNin83}, which involves the study of the evolution of the energy density of charged particles in parallel electric and magnetic fields. We examined this  energy density behavior using two approaches.

First, we have relied on
the well-known classical electrodynamics expression for the energy
change of a charged particle in an external electromagnetic field
in Eq.~(\ref{eq:Ekevolclass}). Then, using the quantum energy spectrum of
massless fermions in Eq.~(\ref{eq:En>0}) and in Table~\ref{tab:Epzn=00003D0},
we have derived the expression for the evolution of the kinetic energy density in Eq.~\eqref{eq:rhomm5v5}. Then, the same result was reproduced in Eq.~\eqref{eq:EdotRL} using the analogue of the classical kinetic equation. 

We have mentioned in Sec.~\ref{sec:ADLER} that, besides the particles acceleration by the external electric field, the change of the number densities of massless right and left particles and antiparticles in parallel electric and magnetic fields can contribute to the evolution of the total energy density. This contribution, driven by the Adler anomaly, is comparable with that found in Sec.~\ref{sec:EVKINEN}. Thus, to reach the agreement between the two approaches for the derivation of CME, proposed in Refs.~\cite{Vil80} and~\cite{NieNin83}, even at the absence of the electroweak background matter, in Sec.~\ref{sec:COMP} we have suggested that one has to omit the contribution of the Adler anomaly to the evolution of the total energy density while using the method developed in Ref.~\cite{NieNin83}.

The Adler anomaly is known to be an essentially quantum phenomenon which can be obtained, e.g., by the calculation of the Feynman loop diagrams~\cite{PesSch95}. Hence it is inexpedient to take it into account in the derivation of CME which turns out to be a quasiclassical effect.

While applying the method in Ref.~\cite{NieNin83} for the description of CME in the presence on the electroweak background method, as well as disregarding the Adler anomaly, as suggested in Sec.~\ref{sec:COMP}, we have got the expression for the anomalous current along the magnetic field in Eq.~\eqref{eq:J5DvoSem}. This expression exactly coincides with that previously obtained in Refs.~\cite{DvoSem15,DvoSem15b}.

Now, it is necessary to analyze the recent derivation of CME in the presence of the electroweak matter made in Ref.~\cite{Kap16}. It was claimed in Ref.~\cite{Kap16} that the analogue of the current in Eq.~\eqref{eq:J5DvoSem} has the form, $\mathbf{J} =  e^{2}\mu_{5}\mathbf{B} / 2\pi^{2}$, i.e. the electroweak interaction would not contribute to $\mathbf{J}$. The method analogous to that in Eqs.~\eqref{eq:Ekchange}-\eqref{eq:EdotRL} was used in Ref.~\cite{Kap16} to obtain this result. Nevertheless, some conceptual errors were made in Ref.~\cite{Kap16}.

Firstly, assuming that only $p_z$ changes when a particle is accelerated in an external electric field $\mathbf{E} = \mathcal{E} \mathbf{e}_z$, i.e. only its kinetic energy changes, the authors of Ref.~\cite{Kap16} studied the change of the total energy in the analogue of Eq.~\eqref{eq:Ekchange}. It is clear that the potential energy of right and left particles and antiparticles, which are equal to $\pm V_\mathrm{R,L}$, cannot change in an external electric field since the number density of background matter is constant by definition.

Secondly, the authors of Ref.~\cite{Kap16} have overlooked that CME is the polarization phenomenon for massless particles at the zero Landau level. As demonstrated in Appendix~\ref{sec:QUAL}, CME should disappear if the lowest energy level is equally populated by right and left particles, i.e. when $n_\mathrm{R}^{(n=0)} = n_\mathrm{L}^{(n=0)}$, cf. Eq.~\eqref{eq:Jz}. In the case when, besides the external magnetic field, massless particles can interact electroweakly with the background matter, this condition is equivalent to $\mu_\mathrm{R} - V_\mathrm{R} = \mu_\mathrm{L} - V_\mathrm{L}$ or $\mu_5 + V_5 = 0$, as results from Eq.~\eqref{eq:npm}. However, the anomalous current derived in Ref.~\cite{Kap16} does not vanish when $\mu_5 = - V_5$.

Finally we mention that the presence of the constant term like $V_5$ in the anomalous electric current in Eq.~\eqref{eq:J5DvoSem} is known not only in the elementary particle physics. Analogous phenomenon also exists in the solid state physics when CME is implemented in Weyl semimetals~\cite{Cor16}. Thus we can conclude that the derivation of CME in the presence of the electroweak background matter made in Ref.~\cite{Kap16} cannot be considered as correct.

Summarizing, we have analyzed the scope of the method for the description of CME proposed in Ref.~\cite{NieNin83}. Neglecting the Adler anomaly, this method gives the results identical to those in Ref.~\cite{Vil80}. Applying the method in Ref.~\cite{NieNin83} to describe CME in the presence of the electroweak background matter, we get the expression for the anomalous current coinciding with that in Refs.~\cite{DvoSem15,DvoSem15b}. We have also compared our results with the findings of other authors.

%

\section*{Acknowledgements}

One of the authors (MD) is thankful to the Competitiveness Improvement Program at the Tomsk State University and
RFBR (research project No.~15-02-00293) for a partial support.

\appendix

\section{Qualitative derivation of CME\label{sec:QUAL}}

It is remarkable that CME can be derived using quite obvious arguments even without the knowledge of the particle wave function $\psi$, which is the solution of the Dirac Eq.~\eqref{eq:Direq}. Below we provide this derivation since it highlights the necessity of the additional term $\sim V_5$ in Eq.~\eqref{eq:J5DvoSem} in the presence of the electroweak background matter.

Massless charged particles interacting with an external magnetic field at the lowest Landau level were mentioned in Ref.~\cite{Vil80} to have the asymmetry in their motion with respect to the magnetic field $\mathbf{B} = \mathcal{B}\mathbf{e}_z$, with $\mathcal{B}>0$. Indeed, if we consider a negatively charged left particle at the zero Landau level, its spin is directed against the magnetic field. The momentum $p_z$, in its turn, is against the particle spin since it is a left particle. Hence $p_z$ should be along $\mathbf{B}$, i.e. $p_z > 0$. Analogously one can conclude that $p_z < 0$ for a right negatively charged particle. This feature remains valid in the electroweak background matter~\cite{DvoSem15,DvoSem15b}, as shown in Table~\ref{tab:Epzn=00003D0}. Note that particles at higher Landau levels with $n>0$ can move arbitrarily with respect to the magnetic field, i.e. $-\infty < p_z < +\infty$ for them.

Let us assume, just for simplicity, that no antiparticles are present
in the system. Taking into account that right and left fermions move in opposite directions with respect
to the magnetic field, one gets that a nonzero current along $\mathbf{B}$, i.e. $J_z$,
can exist only if the zero Landau level is differently populated by left and
right particles, i.e.
\begin{equation}\label{eq:Jz}
  J_{z} \sim e
  \left[
    n_\mathrm{R}^{(n=0)}-n_\mathrm{L}^{(n=0)}
  \right],
\end{equation}
where we take into account that massless particles have the velocity
$v\sim1$ and their electric charge is negative: $q = -e$.

Using Eq.~\eqref{eq:nLdef}, we obtain that the number densities of particles at the zero Landau level $n_\mathrm{R,L}^{(n=0)}$ have that
form,
\begin{align}\label{eq:n0}
  n_\mathrm{R}^{(n=0)} & =
  \frac{e\mathcal{B}}{4\pi^{2}}\int_{-\infty}^{0}\mathrm{d}p_{z}
  f(-p_{z}+V_\mathrm{R}-\mu_\mathrm{R}),
  \nonumber
  \\
  n_\mathrm{L}^{(n=0)} & =
  \frac{e\mathcal{B}}{4\pi^{2}}\int_{0}^{+\infty}\mathrm{d}p_{z}
  f(p_{z}+V_\mathrm{L}-\mu_\mathrm{L}).
\end{align}
For the definiteness, we shall suppose that the electron gas is strongly degenerate, i.e., for
example, $f(p_{z} + V_\mathrm{L} - \mu_\mathrm{L}) = \theta(\mu_\mathrm{L} - V_\mathrm{L} - p_{z})$ etc. In this situation, basing on Eq.~\eqref{eq:n0}, we obtain that
\begin{equation}\label{eq:npm}
  n_\mathrm{R,L}^{(n=0)} =
  \frac{e\mathcal{B}}{4\pi^{2}}(\mu_\mathrm{R,L}-V_\mathrm{R,L}).
\end{equation}
Using Eq.~(\ref{eq:npm}), we can transform Eq.~(\ref{eq:Jz}) to the form,
\begin{equation}\label{eq:Jzfin}
  J_{z} \sim \frac{e^{2}}{2\pi^{2}}
  \left(
    \mu_5+ V_{5}
  \right) \mathcal{B},
\end{equation}
which exactly coincides with the current in Eq.~\eqref{eq:J5DvoSem}.

Unfortunately, this qualitative derivation does not allow one to obtain the current at the arbitrary temperature $T$ of the fermion gas. Nevertheless, the feature that $J_z$ in Eq.~\eqref{eq:Jzfin} should contain $V_\mathrm{R,L}$ remains valid at any $T$ since $n_\mathrm{R,L}^{(n=0)} = n_\mathrm{R,L}^{(n=0)}(\mu_\mathrm{R,L} - V_\mathrm{R,L})$, as results from Eq.~\eqref{eq:n0}.  Moreover, one can see in Eqs.~\eqref{eq:npm} and~(\ref{eq:Jzfin}) that $J_{z}$  is vanishing
if $n_\mathrm{R}^{(n=0)} = n_\mathrm{L}^{(n=0)}$, as it should be. Indeed, in this case, right and left particles form equal electric currents: $|J_{z\mathrm{R}}|=|J_{z\mathrm{L}}|$. However, these currents flow in opposite directions, giving one the
vanishing total current: $J_{z}=J_{z\mathrm{R}}+J_{z\mathrm{L}}=0$.

Finally we mention that, as seen in the above derivation, the asymmetric motion of massless charged particles at the lowest Landau level with respect to the external magnetic field is the key condition for the existence of CME. This feature remains valid if the electroweak background matter is present in the system. If charged particles have any small but nonzero masses, CME does not exist any longer~\cite{Dvo16c}.


\begin{thebibliography}{50}

\bibitem{Kha15}
  {V.A.~Miransky, I.A.~Shovkovy,
  Quantum field theory in a magnetic field:
  From quantum chromodynamics to graphene and Dirac semimetals,
  Phys. Rept. 576 (2015) 1--209}.
  arXiv:1503.00732.

\bibitem{Vil80}
  A.~Vilenkin,
  Equilibrium parity-violating current in a magnetic field,
  Phys. Rev. D 22 (1980) 3080--3084.

\bibitem{NieNin83}
  H.B.~Nielsen, M.~Ninomiya,
  The Adler-Bell-Jackiw anomaly and Weyl fermions in a crystal,
  Phys. Lett. B 130 (1983) 389--396.

\bibitem{DvoSem15}
  M.~Dvornikov, V.B.~Semikoz,
  Magnetic field instability in a neutron star driven
  by the electroweak electron-nucleon interaction
  versus the chiral magnetic effect,
  Phys. Rev. D 91 (2015) 061301,
  arXiv:1410.6676.

\bibitem{DvoSem15b}
  M.~Dvornikov, V.B.~Semikoz,
  Generation of the magnetic helicity in a neutron star driven
  by the electroweak electron-nucleon interaction,
  J. Cosmol. Astropart. Phys. 05 (2015) 032,
  arXiv:1503.04162.

\bibitem{Dvo16}
  M.~Dvornikov,
  Generation of strong magnetic fields in dense quark matter driven
  by the electroweak interaction of quarks,
  Nucl. Phys. B 913 (2016) 79--92,
  arXiv:1608.04946.

%
%

\bibitem{BoyRucSha12}
  {A.~Boyarsky, O.~Ruchayskiy, M.~Shaposhnikov,
  Long-range magnetic fields in the ground state
  of the Standard Model plasma,
  Phys. Rev. Lett. 109 (2012) 111602}.
  arXiv:1204.3604.

\bibitem{Dvo15}
  M.~Dvornikov,
  Galvano-rotational effect induced by electroweak interactions in pulsars,
  J. Cosmol. Astropart. Phys. 05 (2015) 037,
  arXiv:1503.00608.

\bibitem{LanLif88}
  L.D.~Landau, E.M.~Lifshitz,
  The Classical Theory of Fields, 4th ed.,
  Butterworth Heinemann, Amsterdam, 1994.

\bibitem{LanLif02}
  L.D.~Landau, E.M.~Lifshitz,
  Statistical Physics. Part~I, 3rd ed.,
  Pergamon, Oxford, 1980.

\bibitem{Coh77}
  C.~Cohen-Tannoudji, B.~Diu, F.~Lalo\"e,
  Quantum Mechanics. Vol.~1,
  Wiley, New York, 1977, pp.~763--764.

\bibitem{PesSch95}
  M.E.~Peskin, D.V.~Schr\"oder,
  An Introduction to Quantum Field Theory,
  Perseus Books, Reading, MA, 1995, pp.~659--667.

\bibitem{Nun97}
  H.~Nunokawa, V.B.~Semikoz, A.Yu.~Smirnov, J.W.F.~Valle,
  Neutrino conversions in a polarized medium,
  Nucl. Phys. B 501 (1997) 17--40,
  hep-ph/9701420.

\bibitem{Kap16}
  D.B.~Kaplan, S.~Reddy, S.~Sen,
  Energy conservation and the chiral magnetic effect (2016),
  arXiv:1612.00032.




\bibitem{Cor16}
  A.~Cortijo, D.~Kharzeev, K.~Landsteiner, M.A.H.~Vozmediano,
  Strain induced chiral magnetic effect in Weyl semimetals,
  Phys. Rev. B 94 (2016) 241405,
  arXiv:1607.03491.

\bibitem{Dvo16c}
  M.~Dvornikov,
  Role of particle masses in the magnetic field generation driven
  by the parity violating interaction,
  Phys. Lett. B 760 (2016) 406--410,
  arXiv:1608.04940.

\end{thebibliography}
\end{document}